\documentclass[
preprint,
showpacs,preprintnumbers,
bibnotes,
 amsmath,amssymb,
 aps,
 pra,
superscriptaddress,
longbibliography,
]{revtex4-2}

\usepackage[english]{babel}
\usepackage[utf8]{inputenc}
\usepackage{geometry}
\usepackage{amsmath,amsthm,amsfonts,amssymb,amscd,mathtools}
\usepackage{hyperref}
\usepackage{lipsum}
\usepackage{dcolumn}
\usepackage{bm}
\usepackage{enumerate}

\usepackage{textcomp}
\usepackage{graphicx}
\usepackage{fixmath}
\usepackage{xcolor}

\def\beq{\begin{equation}}
\def\eeq{\end{equation}}

\newcommand{\fref}[1]{Fig.~\ref{#1}}

\newcommand{\cref}[1]{chapter~\ref{#1}}

\newcommand{\Cref}[1]{Chapter~\ref{#1}}

\def\3rd{3$^{\text{rd}}$}
\def\5th{5$^{\text{th}}$}

\begin{document}

\title{Spectral tuning of high-harmonic generation with resonance-gradient metasurfaces}

\author{Piyush Jangid}
\altaffiliation{Contributed equally}
\affiliation{Nonlinear Physics Center, Research School of Physics, Australian National University, Canberra ACT 2601, Australia}
\author{Felix Ulrich Richter}
\altaffiliation{Contributed equally}
\affiliation{Institute of Bioengineering, \'Ecole Polytechnique F\'ed\'erale de Lausanne (EPFL), Lausanne 1015, Switzerland}
\author{Ming Lun Tseng}
\affiliation{Institute of Electronics, National Yang Ming Chiao Tung University, Hsinchu 300, Taiwan}
\author{Ivan Sinev}
\affiliation{Institute of Bioengineering, \'Ecole Polytechnique F\'ed\'erale de Lausanne (EPFL), Lausanne 1015, Switzerland}
\author{Sergey Kruk}
\email{Corresponding author: sergey.kruk@anu.edu.au}
\affiliation{Nonlinear Physics Center, Research School of Physics, Australian National University, Canberra ACT 2601, Australia}
\author{Hatice Altug}
\email{Corresponding author: hatice.altug@epfl.ch}
\affiliation{Institute of Bioengineering, \'Ecole Polytechnique F\'ed\'erale de Lausanne (EPFL), Lausanne 1015, Switzerland}
\author{Yuri Kivshar}
\email{Corresponding author: yuri.kivshar@anu.edu.au}
\affiliation{Nonlinear Physics Center, Research School of Physics, Australian National University, Canberra ACT 2601, Australia}



\begin{abstract}
High-index dielectric subwavelength structures and metasurfaces are capable of enhancing light-matter interaction by orders of magnitude via geometry-dependent optical resonances. This enhancement, however, comes with a fundamental limitation of a narrow spectral range of operation in the vicinity of one or few resonant frequencies. Here we tackle this limitation and introduce an innovative and practical approach to achieve spectrally tunable enhancement of light-matter interaction with resonant metasurfaces. We design and fabricate {\it resonance-gradient metasurfaces} with varying geometrical parameters that translate into resonant frequencies dependence on one of the coordinates of the metasurface. The metasurfaces are composed of bone-like nanoresonators which are made of germanium, and they support high-$Q$ optical resonances in the mid-IR spectral range. We apply this general concept to observe the resonant enhancement of the \3rd and \5th harmonics generated from the gradient metasurfaces being used in conjunction with a tunable excitation laser to provide a wide spectral coverage of resonantly-enhanced tunable generation of multiple optical harmonics.
\end{abstract}

\maketitle


\section{\label{sec:intro} Introduction}

Resonant dielectric nanostructures and metasurfaces emerged recently as a new platform to enhance light-matter interaction~\cite{mie2022}. All-dielectric photonic structures featuring localized and extended resonances underpin a rapid development of nonlinear optics at the nanoscale~\cite{Kruk2017,Li2017,Krasnok2017,Dombi2020,ref:Zubyuk2021_strongfield,polina2023}.  This field has progressed from the studies of lower-order, perturbative nonlinear effects such as second-harmonic generation~ (see, e.g., Refs.~\cite{Schlickriede2018,Koshelev2020,leo2022}) and third-harmonics generation~(see, e.g., Refs.~\cite{Shcherbakov2014,Yang2015,Grinblat2016,Kruk2022NPHOT}) towards the studies of nonperturbative nonlinear regimes of light-matter interaction including multi-photon absorption \cite{ref:zalogina2022_photoluminescence} and high-harmonic generation~\cite{Golde2008,Liu2018,Ghimire2019,Shcherbakov2021,Zograf2022,Zalogina2023}.

Nanoparticles with carefully engineered geometries can support optical modes empowered by Mie resonances, such as optically-induced magnetic dipole resonances~\cite{Shcherbakov2014}, higher-order multipoles \cite{Liza2018}, and anapole states~\cite{Grinblat2016}, as well as extended lattice modes associated with Fano resonances \cite{Yang2015} and bound states in the continuum~\cite{Koshelev2020}. High field enhancements near the resonant modes increase the efficiency of nonlinear processes by orders of magnitude. In a striking contrast to nonlinear light–matter interactions in bulk media governed by the phase-matching requirements, nonlinearities of subwavelength resonators are governed by localized modes. Nonlinear light–matter interactions in such systems depend on the field confinement characterized by the quality factor (or the $Q$-factor) of the resonance, an overlap of the modes at the frequencies involved in the nonlinear process~\cite{Celebrano2015} and selection rules imposed by a crystalline orientation of the material~\cite{frizyuk2019second}, as well as a balance of radiative and nonradiative losses~\cite{Koshelev2019}.


One of the challenges in utilizing resonant dielectric nanostructures for strong-field nonlinear optics is their spectrally narrow response limited to the immediate vicinity of one or few resonant frequencies. In order to cover a broader spectral range in experiment or match precisely the resonant wavelength of the nanostructure with a desired excitation or detection wavelength, pixelated designs of multiple individual metasurfaces have been proposed \cite{Tittl2018}. Alternatively, incident angle dependence of the resonance position was demonstrated as a possibility to tune the spectral response of metasurfaces \cite{Leitis2019}.

\begin{figure}
    \centering
    \includegraphics[width=0.6\textwidth]{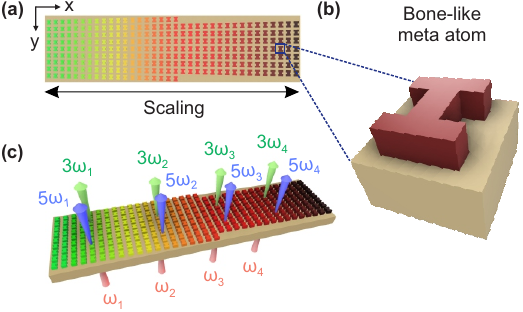}
    \caption{\textbf{Concept of the resonance-gradient metasurface.} (a) Schematic of the gradient metasurface formed by introducing variation of the geometrical parameters along the axes. (b) Metasurface is comprised of a bone-like nanoresonator inside a rectangular lattice. (c) Varying excitation wavelength aligned with varying lateral coordinates of the metasurface produces enhanced and tunable generation of the \3rd and \5th harmonics.}
    \label{fig:concept}
\end{figure}

Here, we tackle this challenge by suggesting and demonstrating an innovative concept of {\it resonance-gradient metasurfaces}. We design, fabricate and characterize experimentally the ``rainbow" dielectric metasurfaces hosting high-$Q$ local Mie-type resonances. By design, the spectral position of the resonance varies continuously with one of the metasurface coordinates. This is achieved by slow and gradual change of the geometric parameters of the constituent elements of the metasurface (see Fig.~1a). 

A gradual change of the structure parameters associated with a resonant frequency shift has been employed earlier in GHz spectral range~\cite{shi2019} in the context of the ``trapped rainbow'' functionality~\cite{boardman2007}, in THz range for imaging applications~\cite{choi2015}, as well as in optical plasmonic systems, in particular for achieving efficient nanofocusing in tapered plasmonic waveguides~\cite{stockman2004,davoyan2010}.  Axially varying metamaterial heterostructures have also been suggested for the concept of ``trapped rainbow" to efficiently and coherently bring light to a complete standstill~\cite{boardman2007}. Nonresonant tapering of the metamaterial waveguides is one of the efficient approaches that allows creating light spots much smaller than the operating wavelength. 

In this paper, we suggest employing gradient metasurfaces for spectrally tunable and resonantly enhanced ``rainbow" generation of high-order optical harmonics upon the excitation of a metasurface with a focused laser beam, as shown schematically in Fig.~1c. More specifically, we design the gradient of metasurface's parameters to be sufficiently slow so that the localized excitation spot of a laser at any given location on the metasurface interacts with the close-to-homogeneous layout of nanoresonators. As we change the coordinate of the excitation spot in accord with the change of the excitation wavelength, we achieve resonantly enhanced and spectrally tunable generation of the \3rd and \5th optical harmonics.

\section{\label{sec:theory} Metasurface design}

\begin{figure*}
    \centering
    \includegraphics[width=\textwidth]{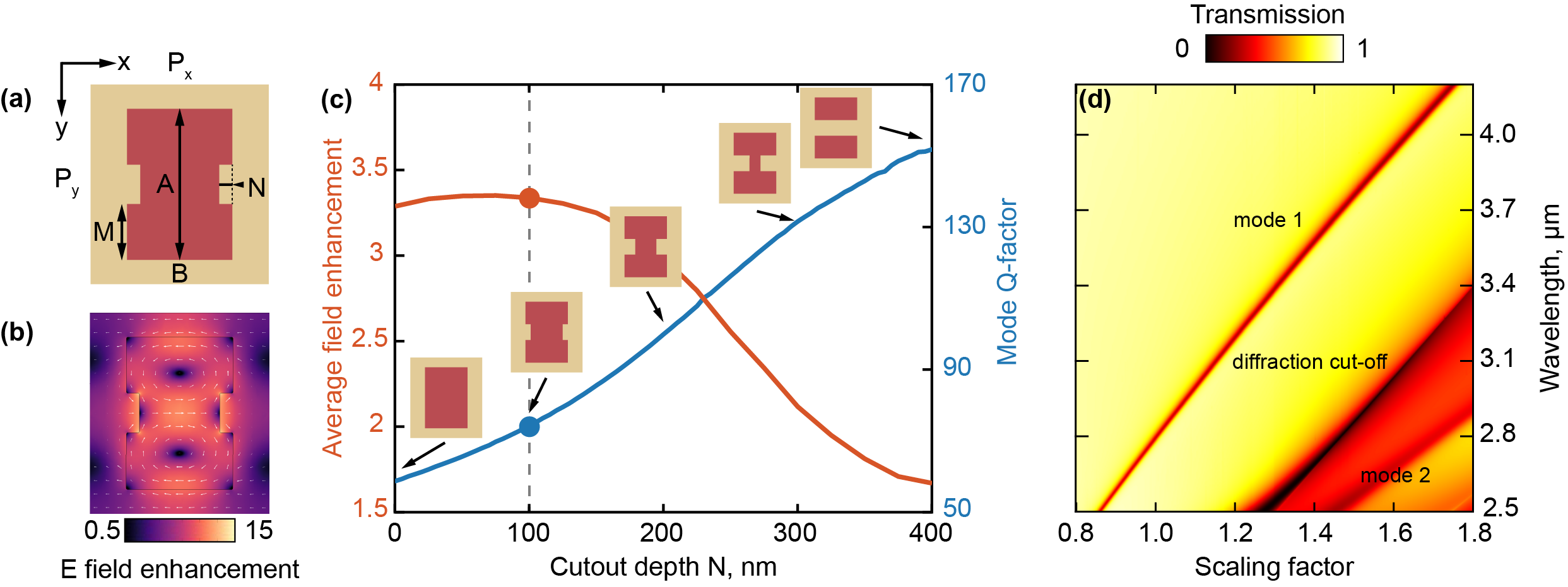}
    \caption{\textbf{Theoretical design of metasurfaces.} (a) Top view of the metasurface unit cell with notations for characteristic dimensions. (b) Map of the resonant electric field enhancement in the unit cell for linearly polarized plane wave excitation (polarization along the horizontal x-axis). The section is taken through the center of the resonator. (c) Dependencies of the resonant mode $Q$-factor (blue line) and the average field enhancement in the resonator volume (red line) on the cutout value. The design chosen for the experimental studies is marked with a dashed line and solid dots. (d) Map of the metasurface transmission spectra for linearly polarized light at normal incidence for different design scaling factors showcasing the gradient concept.}
    \label{fig:analysis}
\end{figure*}

We design the metasurface and optimize its geometric parameters using commercial FEM-based COMSOL Multiphysics. The unit cell design, which is illustrated in detail in Fig.~\ref{fig:analysis}a, comprises a germanium (Ge) bone-like resonator on top of a calcium difluoride (CaF$_2$) substrate. The advantage of the proposed design is that the $Q$-factor of the resonant mode of the metasurface (see Fig.~\ref{fig:analysis}b for its field distribution and Supplementary Materials for a detailed mode analysis) can be tuned by changing the cutout depth N. This allows to precisely match the radiative losses of the mode with fabrication-limited effective nonradiative losses to achieve critical coupling condition and, thus, maximum average local field enhancement, which is beneficial for nonlinear processes. We start with a design scaling factor of 1, which corresponds to P$_x$=1355~nm, P$_y$=1520~nm, M=425~nm, B=800~nm and Ge thickness of 300~nm.  

We estimate the effective non-radiative losses of our metasurface by fitting the measured transmission spectra (see Supplementary Figure S2) and plot the resulting dependence of the average field enhancement in Ge and the mode $Q$-factor in Fig.~\ref{fig:analysis}c. The maximum field enhancement is observed around the cutout value of N=100~nm (corresponding to $Q \approx 75$, marked with solid dots and a dashed line in Fig.~\ref{fig:analysis}c), which is therefore used as a basic design for the gradient dielectric metasurface.

To facilitate the tunable harmonic generation, we scale the chosen design by changing proportionally the in-plane dimensions of the unit cell, including the periodicity. This leads to an almost linear shift of the resonant wavelength with a very minor change in the $Q$-factor as illustrated on the reflectivity map in Fig.~\ref{fig:analysis}d. We choose the scaling factor range (0.8-2) to match the tuning range of the main mode of the Ge resonator with the available spectral bandwidth of the tunable laser (2.5-4.5~$\mu$m). The reflectivity map in Fig.~\ref{fig:analysis}d also shows two other peaks within this range that correspond to the diffraction cut-off and a higher-order secondary mode with similar field distribution (see Supplementary Materials for details).

\section{\label{sec:exp} Results and discussion }

The metasurfaces are fabricated on a CaF$_2$ substrate by patterning a Ge thin film with a single electron beam lithography step (see Supplementary Material). \fref{fig:sem_spectra}a (top and middle) show the optical images of the fabricated gradient metasurface, and \fref{fig:sem_spectra}a (bottom) shows an SEM image of bone-like nanostructures. The chosen materials and critical dimensions of the resonators are compatible with state-of-the-art large-scale fabrication methods such as deep UV lithography or nanoimprint lithography. \cite{Leitis2021}\\

\begin{figure*}
    \centering
    \includegraphics[width=\textwidth]{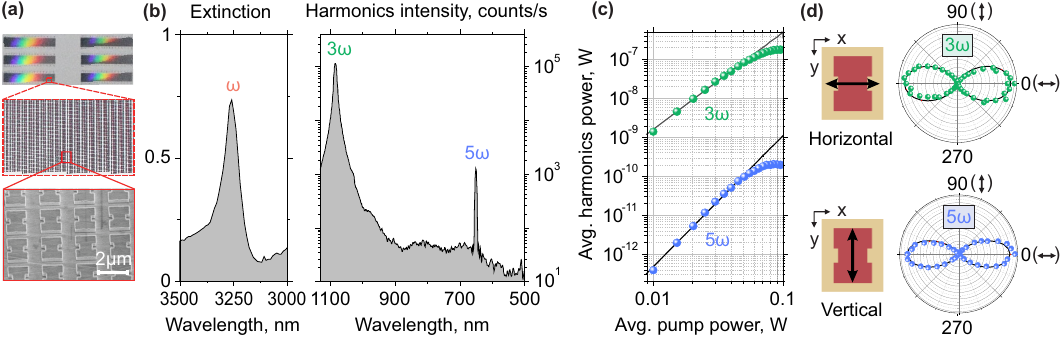}
     \caption{\textbf{Experimental characterisation of metasurfaces.}
     (a) (top) Photo of a nanofabricated chip with 2~mm-long gradient metasurfaces, (middle) optical microscope and (bottom) electron microscope images of bone-like nanostructures of the metasurface. (b) Experimental spectra of the metasurface linear response (left) and nonlinear response (right) featuring a fundamental resonance at $\omega$ and its optical harmonics $3\omega$ and $5\omega$ for the cutout depth N=100~nm. (c) Average power of the \3rd and \5th harmonics versus the average incident power (see Supplementary Material). Logarithmic fitted lines indicate polynomial power dependencies at lower power as: \3rd harmonics power~$\propto$~(Pump~power)$^{2.5}$ and \5th harmonics power~$\propto$~(Pump~power)$^{3.4}$. (d) Dependency of generated \3rd (top) and \5th (bottom) harmonics on the incident polarization shows a cosine curve fit of degrees 3 and 5, respectively. The horizontal and vertical polarization directions inside the unit cell are shown.}
    \label{fig:sem_spectra}
\end{figure*}

To characterize the resonance properties of the metasurface, we perform linear spectroscopy of the metasurface using a Bruker FTIR spectrometer. Light generated by a broadband glow bar is loosely focused onto the metasurface by a low NA refractive zinc selenide (ZnSe) lens. The transmitted light is collected by a custom-built ZnSe objective and focused onto a MCT focal plane array (FPA) detector with $64 \times 64$ pixels. Each pixel column is therefore conjugated with a low-width rectangular part of the metasurface featuring a narrow resonance at a specific wavelength. 
\fref{fig:hhg_scans}a shows the map of measured transmission spectra of small segments of the metasurface defined by the pixel columns of the FPA. The gradual change of the resonance dip with respect to the lateral position on the metasurface is clearly observed and is in perfect agreement with the simulated data in Fig.~\ref{fig:analysis}d. \\

\begin{figure*}[t]
    \centering
    \includegraphics[width=\textwidth]{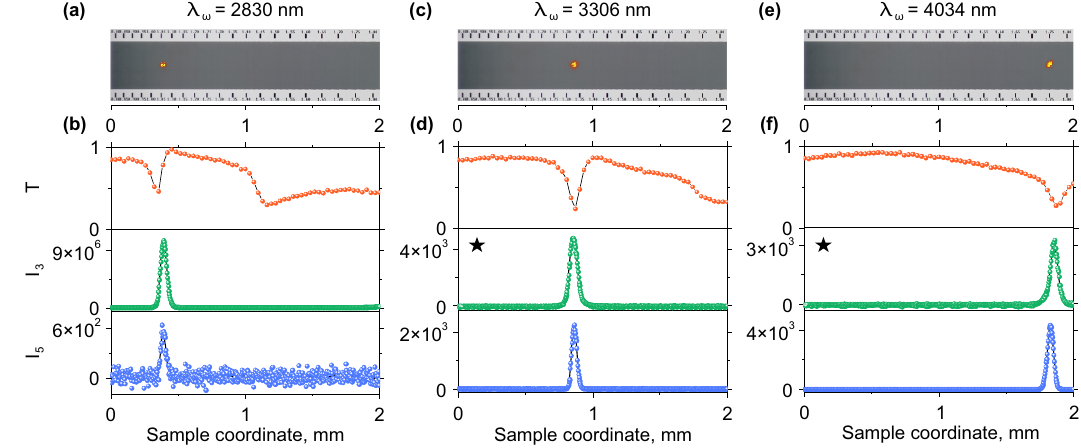}
    \caption{\textbf{Observation of position-dependent resonant response.} (a,c,e) Visible CCD camera images of the harmonics generated at distinct highly resonant metasurface coordinates when pumped at wavelength $\lambda_{\omega}$~=~2830~nm, 3306~nm and 4034~nm, respectively. (b,d,f) Corresponding linear transmission ($T$), intensity of \3rd ($I_{3}$) and \5th ($I_{5}$) harmonics in counts/s further normalized by the quantum efficiency of the spectrometers' detectors, from the top to the bottom, along the metasurface lateral coordinate. $\star$ marked \3rd harmonics spectra are obtained via InGaAs-based spectrometer and the rest \3rd and \5th harmonics spectra via Si-based spectrometer.}
    \label{fig:spots}
\end{figure*}

In our nonlinear experiments, we excite the metasurface with a tunable pulsed laser system consisting of an optical parametric amplifier MIROPA by Hotlight Systems producing idler wavelength tunable in the range 2.5-4.5~$\mu$m pumped by a pulsed laser Ekspla by Femtolux with 1030~nm wavelength. We work with pulses of 10-100~mW range of average power, 3.99 ps pulse duration and 5.14 MHz repetition rate. As we tune the excitation, we maintain a constant level of the average power to enable an unambiguous comparison of the harmonics intensities at different wavelengths. The collimated laser beam with a diameter slightly larger than 4 mm is focused on the metasurface by a CaF$_2$ lens f=40~mm, resulting in a focal spot of approximately 68~$\mu$m diameter at 3.25~$\mu$m incident wavelength and a peak power density of about 12.5-125~MW/cm$^2$. We image the mid-IR excitation spot on the metasurface in reflection captured by the same lens and sent by a beam splitter onto a camera NIT Tachyon 16k paired with a CaF$_2$ lens f=250~mm. We collect the generated harmonics in transmission with an objective lens (Mitutoyo Plan Apo NIR 20$\times$ 0.4NA) and detect them with a Peltier-cooled spectrometers Ocean Optics QE Pro (Silicon-based detector with the detection range: 0.35-1.1~$\mu$m) and NIR Quest (InGaAs-based detector with the detection range: 0.9-1.7~$\mu$m). To estimate the harmonics' irradiance from the spectrometers, we take into account the quantum efficiencies of the spectrometers' detectors and the diffraction efficiencies of their gratings.

In \fref{fig:sem_spectra}b (left), we show a representative spectrum of the detected \3rd ($3\omega$) and \5th ($5\omega$) harmonics. Here the metasurface is exemplary excited with the pump wavelength $\lambda_{\omega} = 3250$~nm (see \fref{fig:sem_spectra}b, right). The maximum of the generated harmonics intensity occur when the pump is horizontally polarized (consistent with the resonant mode polarization), and vanishes for the corresponding orthogonal polarization, \fref{fig:sem_spectra}d. We empirically find that the harmonics intensity versus the incident polarization angle $\phi$ is close to $cos(\phi)^3$ for the third harmonic and close to $cos(\phi)^5$ for the fifth harmonic. We further study the dependence of the \3rd and \5th harmonics power on the average incident power as shown in \fref{fig:sem_spectra}c. At lower levels of the incident power, the harmonics generation follows the polynomial power dependencies, albeit inconsistent with the scaling laws of perturbative nonlinear optics. At higher levels of the incident power, the generation process deviates from polynomial power dependencies, which indicates a possible onset of pump-induced refractive index changes\cite{Sinev2021observation}.

Next, we study the dependence of the harmonics signal versus the pump wavelength and versus the metasurface lateral coordinate, in which we maintain 46~mW of average power for all incident pump wavelengths. First, we illustratively excite the metasurface with $\lambda_{\omega}=$ 2830~nm, 3306~nm and 4034~nm, and record photographs of the sample areas glowing with harmonics on a visible CCD camera (Starlight Xpress SX-694), which are shown to the scale on the metasurface in a false-color on top of grayscale microphotographs of the metasurface (see \fref{fig:spots}a,c and e, respectively). In \fref{fig:spots}b,d and f, the variation of \3rd and \5th harmonics generation intensities with coordinate at $\lambda_{\omega}=$ 2830~nm, 3306~nm and 4034~nm, respectively, are plotted in the middle and bottom panels. They appear precisely at the resonance location in the corresponding linear spectrum shown in the top panel. We observe a sharp resonant enhancement of the harmonics generation once the incident wavelength is aligned with the corresponding resonating area of the metasurface.

\begin{figure*}[t]
    \centering
    \includegraphics[width=\textwidth]{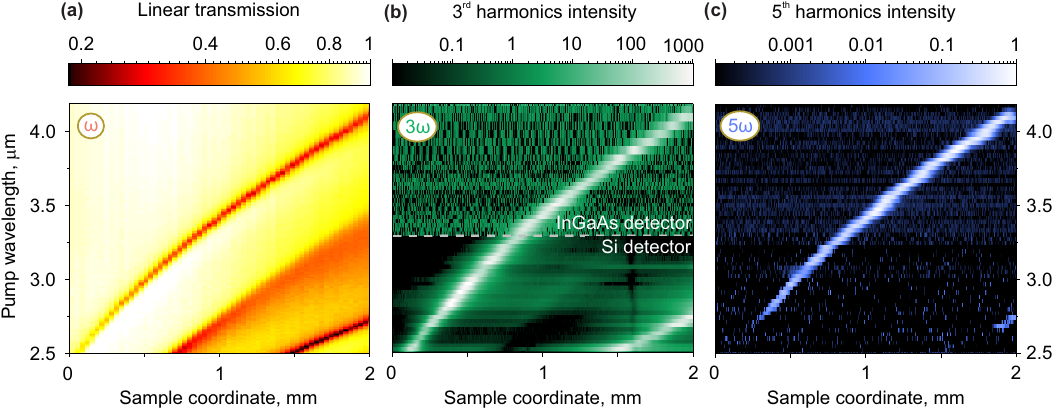}
    \caption{\textbf{Experimental spectral tuning of the harmonic generation.} (a) Map of the linear transmission spectra of the metasurface with the cutout depth N=100~nm showing a continuous change of the resonant wavelength with coordinate along the 2~mm strip. The linear transmission is on a logarithmic scale. (b,c) Continuous tuning of resonant \3rd and \5th harmonics generation, respectively, by simultaneous change of the excitation wavelength and the position of the pump along the gradient metasurface. The \3rd and \5th harmonics intensities, normalized to the maximum of \5th harmonics intensity and quantum efficiencies of the detectors, are in logarithmic scale.} 
    \label{fig:hhg_scans}
\end{figure*}

Next, we monitor the harmonics' wavelength with a spectrometer as we perform a two-dimensional experimental parameter scan of the harmonics intensity vs. pump wavelength vs. pump position on the sample. The resulting maps for the third and fifth harmonics are shown in \fref{fig:hhg_scans}b and c, with harmonic wavelengths covering the spectral ranges of 830-1430~nm and 500-860~nm, respectively. We note that the \3rd harmonics data was obtained from a combination of two spectrometers: Si-based and InGaAs-based spectrometers. The data from the two spectrometers was appended by using the overlapping spectral region. Other discrepancies between the two spectrometers were neglected. Both maps show two position-dependent peaks that correspond to the modes observed in the linear transmission map of~\fref{fig:hhg_scans}a. Remarkably, the measured harmonic intensity maintains near-constant value across the broad tuning range of the pump, highlighting robustness of the proposed design for gradient scaling. Resonant enhancement of the local field provides considerable enhancement of the harmonics signal, as for an unstructured Ge film of the same thickness and for the same measurement settings, we did not detect any harmonics.

\section{\label{sec:conclusion} Conclusion}

We have introduced a novel type of metasurface with spatially varying parameters to achieve spectrally tunable enhancement of resonant light-matter interaction. The designed and fabricated gradient metasurfaces are based on Ge subwavelength resonators, and they support high-$Q$ optical resonances in the mid-IR spectral range, which facilitates the efficient generation of higher optical harmonics. The resonant wavelength of our gradient metasurfaces varies continuously with the sample coordinate, allowing to realize a tunable high-harmonics source of a nanoscale thickness. In conjunction with a tunable excitation laser, we have demonstrated the resonant enhancement and spectral tunability of the  \3rd and \5th harmonics generated from the metasurfaces. Wide spectral coverage of resonantly-enhanced generation of optical harmonics endows metasurfaces with functionality that was so far only available for bulk nonlinear crystals and adds another degree of freedom for utilizing strong nonlinear response of resonant subwavelength photonic structures.

\section*{Acknowledgements}

The authors thank Dr. Kirill Koshelev for his help with the modal analysis. This work was supported by the Australian Research Council (grants DE210100679 and DP210101292), National Science Council (grant MOST 111-2636-M-A49-005), Ministry of Education (Yushan Young Scholar Program) and the Center for Integrated Electronics-Optics Technologies and Systems, National Yang Ming Chiao Tung University in Taiwan, International Technology Center Indo-Pacific (ITC IPAC) through the Army Research Office (contract FA520921P0034), the European Innovation Council (grant 10104642) and the European Research Council (grants 682167 VIBRANT-BIO and 875672 POCSEL). The authors acknowledge the use of nanofabrication facilities at the Center of MicroNano Technology of \'Ecole Polytechnique F\'ed\'erale de Lausanne.

\newpage 

\bibliography{bibliography_HHG}

\begin{thebibliography}{34}%
\makeatletter
\providecommand \@ifxundefined [1]{%
 \@ifx{#1\undefined}
}%
\providecommand \@ifnum [1]{%
 \ifnum #1\expandafter \@firstoftwo
 \else \expandafter \@secondoftwo
 \fi
}%
\providecommand \@ifx [1]{%
 \ifx #1\expandafter \@firstoftwo
 \else \expandafter \@secondoftwo
 \fi
}%
\providecommand \natexlab [1]{#1}%
\providecommand \enquote  [1]{``#1''}%
\providecommand \bibnamefont  [1]{#1}%
\providecommand \bibfnamefont [1]{#1}%
\providecommand \citenamefont [1]{#1}%
\providecommand \href@noop [0]{\@secondoftwo}%
\providecommand \href [0]{\begingroup \@sanitize@url \@href}%
\providecommand \@href[1]{\@@startlink{#1}\@@href}%
\providecommand \@@href[1]{\endgroup#1\@@endlink}%
\providecommand \@sanitize@url [0]{\catcode `\\12\catcode `\$12\catcode
  `\&12\catcode `\#12\catcode `\^12\catcode `\_12\catcode `\%12\relax}%
\providecommand \@@startlink[1]{}%
\providecommand \@@endlink[0]{}%
\providecommand \url  [0]{\begingroup\@sanitize@url \@url }%
\providecommand \@url [1]{\endgroup\@href {#1}{\urlprefix }}%
\providecommand \urlprefix  [0]{URL }%
\providecommand \Eprint [0]{\href }%
\providecommand \doibase [0]{https://doi.org/}%
\providecommand \selectlanguage [0]{\@gobble}%
\providecommand \bibinfo  [0]{\@secondoftwo}%
\providecommand \bibfield  [0]{\@secondoftwo}%
\providecommand \translation [1]{[#1]}%
\providecommand \BibitemOpen [0]{}%
\providecommand \bibitemStop [0]{}%
\providecommand \bibitemNoStop [0]{.\EOS\space}%
\providecommand \EOS [0]{\spacefactor3000\relax}%
\providecommand \BibitemShut  [1]{\csname bibitem#1\endcsname}%
\let\auto@bib@innerbib\@empty
\bibitem [{\citenamefont {Kivshar}(2022)}]{mie2022}%
  \BibitemOpen
  \bibfield  {author} {\bibinfo {author} {\bibfnamefont {Y.}~\bibnamefont
  {Kivshar}},\ }\bibfield  {title} {\bibinfo {title} {The rise of
  \uppercase{M}ie-tronics},\ }\href@noop {} {\bibfield  {journal} {\bibinfo
  {journal} {Nano Letters}\ }\textbf {\bibinfo {volume} {22}},\ \bibinfo
  {pages} {3513} (\bibinfo {year} {2022})}\BibitemShut {NoStop}%
\bibitem [{\citenamefont {Kruk}\ and\ \citenamefont
  {Kivshar}(2017)}]{Kruk2017}%
  \BibitemOpen
  \bibfield  {author} {\bibinfo {author} {\bibfnamefont {S.}~\bibnamefont
  {Kruk}}\ and\ \bibinfo {author} {\bibfnamefont {Y.}~\bibnamefont {Kivshar}},\
  }\bibfield  {title} {\bibinfo {title} {Functional meta-optics and
  nanophotonics governed by \uppercase{M}ie resonances},\ }\href
  {https://doi.org/10.1021/acsphotonics.7b01038} {\bibfield  {journal}
  {\bibinfo  {journal} {ACS Photonics}\ }\textbf {\bibinfo {volume} {4}},\
  \bibinfo {pages} {2638} (\bibinfo {year} {2017})}\BibitemShut {NoStop}%
\bibitem [{\citenamefont {Li}\ \emph {et~al.}(2017)\citenamefont {Li},
  \citenamefont {Zhang},\ and\ \citenamefont {Zentgraf}}]{Li2017}%
  \BibitemOpen
  \bibfield  {author} {\bibinfo {author} {\bibfnamefont {G.}~\bibnamefont
  {Li}}, \bibinfo {author} {\bibfnamefont {S.}~\bibnamefont {Zhang}},\ and\
  \bibinfo {author} {\bibfnamefont {T.}~\bibnamefont {Zentgraf}},\ }\bibfield
  {title} {\bibinfo {title} {Nonlinear photonic metasurfaces},\ }\href
  {https://doi.org/10.1038/natrevmats.2017.10} {\bibfield  {journal} {\bibinfo
  {journal} {Nature Reviews Materials}\ }\textbf {\bibinfo {volume} {2}},\
  \bibinfo {pages} {17010} (\bibinfo {year} {2017})}\BibitemShut {NoStop}%
\bibitem [{\citenamefont {Krasnok}\ \emph {et~al.}(2017)\citenamefont
  {Krasnok}, \citenamefont {Tymchenko},\ and\ \citenamefont
  {Al\'u}}]{Krasnok2017}%
  \BibitemOpen
  \bibfield  {author} {\bibinfo {author} {\bibfnamefont {A.}~\bibnamefont
  {Krasnok}}, \bibinfo {author} {\bibfnamefont {M.}~\bibnamefont {Tymchenko}},\
  and\ \bibinfo {author} {\bibfnamefont {A.}~\bibnamefont {Al\'u}},\ }\bibfield
   {title} {\bibinfo {title} {Nonlinear metasurfaces: \uppercase{a} paradigm
  shift in nonlinear optics},\ }\href@noop {} {\bibfield  {journal} {\bibinfo
  {journal} {Materials Today}\ }\textbf {\bibinfo {volume} {21}},\ \bibinfo
  {pages} {8} (\bibinfo {year} {2017})}\BibitemShut {NoStop}%
\bibitem [{\citenamefont {Dombi}\ \emph {et~al.}(2020)\citenamefont {Dombi},
  \citenamefont {P\'apa}, \citenamefont {Vogelsang}, \citenamefont {Yalunin},
  \citenamefont {Sivis}, \citenamefont {Herink}, \citenamefont {Sch\"afer},
  \citenamefont {Gross}, \citenamefont {Ropers},\ and\ \citenamefont
  {Lienau}}]{Dombi2020}%
  \BibitemOpen
  \bibfield  {author} {\bibinfo {author} {\bibfnamefont {P.}~\bibnamefont
  {Dombi}}, \bibinfo {author} {\bibfnamefont {Z.}~\bibnamefont {P\'apa}},
  \bibinfo {author} {\bibfnamefont {J.}~\bibnamefont {Vogelsang}}, \bibinfo
  {author} {\bibfnamefont {S.}~\bibnamefont {Yalunin}}, \bibinfo {author}
  {\bibfnamefont {M.}~\bibnamefont {Sivis}}, \bibinfo {author} {\bibfnamefont
  {G.}~\bibnamefont {Herink}}, \bibinfo {author} {\bibfnamefont
  {S.}~\bibnamefont {Sch\"afer}}, \bibinfo {author} {\bibfnamefont
  {P.}~\bibnamefont {Gross}}, \bibinfo {author} {\bibfnamefont
  {C.}~\bibnamefont {Ropers}},\ and\ \bibinfo {author} {\bibfnamefont
  {C.}~\bibnamefont {Lienau}},\ }\bibfield  {title} {\bibinfo {title}
  {Strong-field nano-optics},\ }\href
  {https://doi.org/10.1103/REVMODPHYS.92.025003/FIGURES/45/MEDIUM} {\bibfield
  {journal} {\bibinfo  {journal} {Reviews of Modern Physics}\ }\textbf
  {\bibinfo {volume} {92}},\ \bibinfo {pages} {025003} (\bibinfo {year}
  {2020})}\BibitemShut {NoStop}%
\bibitem [{\citenamefont {Zubyuk}\ \emph {et~al.}(2021)\citenamefont {Zubyuk},
  \citenamefont {Carletti}, \citenamefont {Shcherbakov},\ and\ \citenamefont
  {Kruk}}]{ref:Zubyuk2021_strongfield}%
  \BibitemOpen
  \bibfield  {author} {\bibinfo {author} {\bibfnamefont {V.}~\bibnamefont
  {Zubyuk}}, \bibinfo {author} {\bibfnamefont {L.}~\bibnamefont {Carletti}},
  \bibinfo {author} {\bibfnamefont {M.}~\bibnamefont {Shcherbakov}},\ and\
  \bibinfo {author} {\bibfnamefont {S.}~\bibnamefont {Kruk}},\ }\bibfield
  {title} {\bibinfo {title} {Resonant dielectric metasurfaces in strong optical
  fields},\ }\href@noop {} {\bibfield  {journal} {\bibinfo  {journal} {APL
  Materials}\ }\textbf {\bibinfo {volume} {9}},\ \bibinfo {pages} {060701}
  (\bibinfo {year} {2021})}\BibitemShut {NoStop}%
\bibitem [{\citenamefont {Vabishchevich}\ and\ \citenamefont
  {Kivshar}(2023)}]{polina2023}%
  \BibitemOpen
  \bibfield  {author} {\bibinfo {author} {\bibfnamefont {P.}~\bibnamefont
  {Vabishchevich}}\ and\ \bibinfo {author} {\bibfnamefont {Y.}~\bibnamefont
  {Kivshar}},\ }\bibfield  {title} {\bibinfo {title} {Nonlinear photonics with
  metasurfaces},\ }\href@noop {} {\bibfield  {journal} {\bibinfo  {journal}
  {Photonics Research}\ }\textbf {\bibinfo {volume} {11}},\ \bibinfo {pages}
  {B50} (\bibinfo {year} {2023})}\BibitemShut {NoStop}%
\bibitem [{\citenamefont {Schlickriede}\ \emph {et~al.}(2018)\citenamefont
  {Schlickriede}, \citenamefont {Waterman}, \citenamefont {Reineke},
  \citenamefont {Georgi}, \citenamefont {Li}, \citenamefont {Zhang},\ and\
  \citenamefont {Zentgraf}}]{Schlickriede2018}%
  \BibitemOpen
  \bibfield  {author} {\bibinfo {author} {\bibfnamefont {C.}~\bibnamefont
  {Schlickriede}}, \bibinfo {author} {\bibfnamefont {N.}~\bibnamefont
  {Waterman}}, \bibinfo {author} {\bibfnamefont {B.}~\bibnamefont {Reineke}},
  \bibinfo {author} {\bibfnamefont {P.}~\bibnamefont {Georgi}}, \bibinfo
  {author} {\bibfnamefont {G.}~\bibnamefont {Li}}, \bibinfo {author}
  {\bibfnamefont {S.}~\bibnamefont {Zhang}},\ and\ \bibinfo {author}
  {\bibfnamefont {T.}~\bibnamefont {Zentgraf}},\ }\bibfield  {title} {\bibinfo
  {title} {{Imaging through nonlinear metalens using second-harmonic
  generation}},\ }\href {https://doi.org/10.1002/adma.201703843} {\bibfield
  {journal} {\bibinfo  {journal} {Advanced Materials}\ }\textbf {\bibinfo
  {volume} {30}},\ \bibinfo {pages} {1703843} (\bibinfo {year}
  {2018})}\BibitemShut {NoStop}%
\bibitem [{\citenamefont {Koshelev}\ \emph {et~al.}(2020)\citenamefont
  {Koshelev}, \citenamefont {Kruk}, \citenamefont {Melik-Gaykazyan},
  \citenamefont {Choi}, \citenamefont {Bogdanov}, \citenamefont {Park},\ and\
  \citenamefont {Kivshar}}]{Koshelev2020}%
  \BibitemOpen
  \bibfield  {author} {\bibinfo {author} {\bibfnamefont {K.}~\bibnamefont
  {Koshelev}}, \bibinfo {author} {\bibfnamefont {S.}~\bibnamefont {Kruk}},
  \bibinfo {author} {\bibfnamefont {E.}~\bibnamefont {Melik-Gaykazyan}},
  \bibinfo {author} {\bibfnamefont {J.-H.}\ \bibnamefont {Choi}}, \bibinfo
  {author} {\bibfnamefont {A.}~\bibnamefont {Bogdanov}}, \bibinfo {author}
  {\bibfnamefont {H.-G.}\ \bibnamefont {Park}},\ and\ \bibinfo {author}
  {\bibfnamefont {Y.}~\bibnamefont {Kivshar}},\ }\bibfield  {title} {\bibinfo
  {title} {Subwavelength dielectric resonators for nonlinear nanophotonics.},\
  }\href {https://doi.org/10.1126/science.aaz3985} {\bibfield  {journal}
  {\bibinfo  {journal} {Science}\ }\textbf {\bibinfo {volume} {367}},\ \bibinfo
  {pages} {288} (\bibinfo {year} {2020})}\BibitemShut {NoStop}%
\bibitem [{\citenamefont {Gigli}\ and\ \citenamefont {Leo}(2022)}]{leo2022}%
  \BibitemOpen
  \bibfield  {author} {\bibinfo {author} {\bibfnamefont {C.}~\bibnamefont
  {Gigli}}\ and\ \bibinfo {author} {\bibfnamefont {G.}~\bibnamefont {Leo}},\
  }\bibfield  {title} {\bibinfo {title} {All-dielectric $\chi^{(2)}$
  metasurfaces: recent progress},\ }\href@noop {} {\bibfield  {journal}
  {\bibinfo  {journal} {Opto-Electronic Advances}\ }\textbf {\bibinfo {volume}
  {5}},\ \bibinfo {pages} {210093} (\bibinfo {year} {2022})}\BibitemShut
  {NoStop}%
\bibitem [{\citenamefont {Shcherbakov}\ \emph {et~al.}(2014)\citenamefont
  {Shcherbakov}, \citenamefont {Neshev}, \citenamefont {Hopkins}, \citenamefont
  {Shorokhov}, \citenamefont {Staude}, \citenamefont {Melik-Gaykazyan},
  \citenamefont {Decker}, \citenamefont {Ezhov}, \citenamefont
  {Miroshnichenko}, \citenamefont {Brener}, \citenamefont {Fedyanin},\ and\
  \citenamefont {Kivshar}}]{Shcherbakov2014}%
  \BibitemOpen
  \bibfield  {author} {\bibinfo {author} {\bibfnamefont {M.}~\bibnamefont
  {Shcherbakov}}, \bibinfo {author} {\bibfnamefont {D.}~\bibnamefont {Neshev}},
  \bibinfo {author} {\bibfnamefont {B.}~\bibnamefont {Hopkins}}, \bibinfo
  {author} {\bibfnamefont {A.}~\bibnamefont {Shorokhov}}, \bibinfo {author}
  {\bibfnamefont {I.}~\bibnamefont {Staude}}, \bibinfo {author} {\bibfnamefont
  {E.}~\bibnamefont {Melik-Gaykazyan}}, \bibinfo {author} {\bibfnamefont
  {M.}~\bibnamefont {Decker}}, \bibinfo {author} {\bibfnamefont
  {A.}~\bibnamefont {Ezhov}}, \bibinfo {author} {\bibfnamefont
  {A.}~\bibnamefont {Miroshnichenko}}, \bibinfo {author} {\bibfnamefont
  {I.}~\bibnamefont {Brener}}, \bibinfo {author} {\bibfnamefont
  {A.}~\bibnamefont {Fedyanin}},\ and\ \bibinfo {author} {\bibfnamefont
  {Y.}~\bibnamefont {Kivshar}},\ }\bibfield  {title} {\bibinfo {title}
  {Enhanced third-harmonic generation in silicon nanoparticles driven by
  magnetic response},\ }\href {https://doi.org/10.1021/nl503029j} {\bibfield
  {journal} {\bibinfo  {journal} {Nano Letters}\ }\textbf {\bibinfo {volume}
  {14}},\ \bibinfo {pages} {6488} (\bibinfo {year} {2014})}\BibitemShut
  {NoStop}%
\bibitem [{\citenamefont {Yang}\ \emph {et~al.}(2015)\citenamefont {Yang},
  \citenamefont {Wang}, \citenamefont {Boulesbaa}, \citenamefont {Kravchenko},
  \citenamefont {Briggs}, \citenamefont {Puretzky}, \citenamefont {Geohegan},\
  and\ \citenamefont {Valentine}}]{Yang2015}%
  \BibitemOpen
  \bibfield  {author} {\bibinfo {author} {\bibfnamefont {Y.}~\bibnamefont
  {Yang}}, \bibinfo {author} {\bibfnamefont {W.}~\bibnamefont {Wang}}, \bibinfo
  {author} {\bibfnamefont {A.}~\bibnamefont {Boulesbaa}}, \bibinfo {author}
  {\bibfnamefont {I.}~\bibnamefont {Kravchenko}}, \bibinfo {author}
  {\bibfnamefont {D.}~\bibnamefont {Briggs}}, \bibinfo {author} {\bibfnamefont
  {A.}~\bibnamefont {Puretzky}}, \bibinfo {author} {\bibfnamefont
  {D.}~\bibnamefont {Geohegan}},\ and\ \bibinfo {author} {\bibfnamefont
  {J.}~\bibnamefont {Valentine}},\ }\bibfield  {title} {\bibinfo {title}
  {Nonlinear fano-resonant dielectric metasurfaces},\ }\href
  {https://doi.org/10.1021/acs.nanolett.5b02802} {\bibfield  {journal}
  {\bibinfo  {journal} {Nano Letters}\ }\textbf {\bibinfo {volume} {15}},\
  \bibinfo {pages} {7388} (\bibinfo {year} {2015})}\BibitemShut {NoStop}%
\bibitem [{\citenamefont {Grinblat}\ \emph {et~al.}(2016)\citenamefont
  {Grinblat}, \citenamefont {Li}, \citenamefont {Nielsen}, \citenamefont
  {Oulton},\ and\ \citenamefont {Maier}}]{Grinblat2016}%
  \BibitemOpen
  \bibfield  {author} {\bibinfo {author} {\bibfnamefont {G.}~\bibnamefont
  {Grinblat}}, \bibinfo {author} {\bibfnamefont {Y.}~\bibnamefont {Li}},
  \bibinfo {author} {\bibfnamefont {M.}~\bibnamefont {Nielsen}}, \bibinfo
  {author} {\bibfnamefont {R.}~\bibnamefont {Oulton}},\ and\ \bibinfo {author}
  {\bibfnamefont {S.}~\bibnamefont {Maier}},\ }\bibfield  {title} {\bibinfo
  {title} {Enhanced third-harmonic generation in single germanium nanodisks
  excited at the anapole mode},\ }\href
  {https://doi.org/10.1021/acs.nanolett.6b01958} {\bibfield  {journal}
  {\bibinfo  {journal} {Nano Letters}\ }\textbf {\bibinfo {volume} {16}},\
  \bibinfo {pages} {4635} (\bibinfo {year} {2016})}\BibitemShut {NoStop}%
\bibitem [{\citenamefont {Kruk}\ \emph {et~al.}(2022)\citenamefont {Kruk},
  \citenamefont {Wang}, \citenamefont {Sain}, \citenamefont {Dong},
  \citenamefont {Yang}, \citenamefont {Zentgraf},\ and\ \citenamefont
  {Kivshar}}]{Kruk2022NPHOT}%
  \BibitemOpen
  \bibfield  {author} {\bibinfo {author} {\bibfnamefont {S.}~\bibnamefont
  {Kruk}}, \bibinfo {author} {\bibfnamefont {L.}~\bibnamefont {Wang}}, \bibinfo
  {author} {\bibfnamefont {B.}~\bibnamefont {Sain}}, \bibinfo {author}
  {\bibfnamefont {Z.}~\bibnamefont {Dong}}, \bibinfo {author} {\bibfnamefont
  {J.}~\bibnamefont {Yang}}, \bibinfo {author} {\bibfnamefont {T.}~\bibnamefont
  {Zentgraf}},\ and\ \bibinfo {author} {\bibfnamefont {Y.}~\bibnamefont
  {Kivshar}},\ }\bibfield  {title} {\bibinfo {title} {{Asymmetric parametric
  generation of images with nonlinear dielectric metasurfaces}},\ }\href
  {https://doi.org/10.1038/s41566-022-01018-7} {\bibfield  {journal} {\bibinfo
  {journal} {Nature Photonics}\ }\textbf {\bibinfo {volume} {16}},\ \bibinfo
  {pages} {561} (\bibinfo {year} {2022})}\BibitemShut {NoStop}%
\bibitem [{\citenamefont {Zalogina}\ \emph {et~al.}(2022)\citenamefont
  {Zalogina}, \citenamefont {Tonkaev}, \citenamefont {Tripathi}, \citenamefont
  {Lee}, \citenamefont {Carletti}, \citenamefont {Park}, \citenamefont {Kruk},\
  and\ \citenamefont {Kivshar}}]{ref:zalogina2022_photoluminescence}%
  \BibitemOpen
  \bibfield  {author} {\bibinfo {author} {\bibfnamefont {A.}~\bibnamefont
  {Zalogina}}, \bibinfo {author} {\bibfnamefont {P.}~\bibnamefont {Tonkaev}},
  \bibinfo {author} {\bibfnamefont {A.}~\bibnamefont {Tripathi}}, \bibinfo
  {author} {\bibfnamefont {H.-C.}\ \bibnamefont {Lee}}, \bibinfo {author}
  {\bibfnamefont {L.}~\bibnamefont {Carletti}}, \bibinfo {author}
  {\bibfnamefont {H.-G.}\ \bibnamefont {Park}}, \bibinfo {author}
  {\bibfnamefont {S.}~\bibnamefont {Kruk}},\ and\ \bibinfo {author}
  {\bibfnamefont {Y.}~\bibnamefont {Kivshar}},\ }\bibfield  {title} {\bibinfo
  {title} {Enhanced five-photon photoluminescence in subwavelength
  \uppercase{A}l\uppercase{G}a\uppercase{A}s resonators},\ }\href@noop {}
  {\bibfield  {journal} {\bibinfo  {journal} {Nano Letters}\ }\textbf {\bibinfo
  {volume} {22}},\ \bibinfo {pages} {4200} (\bibinfo {year}
  {2022})}\BibitemShut {NoStop}%
\bibitem [{\citenamefont {Golde}\ \emph {et~al.}(2008)\citenamefont {Golde},
  \citenamefont {Meier},\ and\ \citenamefont {Koch}}]{Golde2008}%
  \BibitemOpen
  \bibfield  {author} {\bibinfo {author} {\bibfnamefont {D.}~\bibnamefont
  {Golde}}, \bibinfo {author} {\bibfnamefont {T.}~\bibnamefont {Meier}},\ and\
  \bibinfo {author} {\bibfnamefont {S.}~\bibnamefont {Koch}},\ }\bibfield
  {title} {\bibinfo {title} {High harmonics generated in semiconductor
  nanostructures by the coupled dynamics of optical inter- and intraband
  excitations},\ }\href {https://doi.org/10.1103/PhysRevB.77.075330} {\bibfield
   {journal} {\bibinfo  {journal} {Physical Review B}\ }\textbf {\bibinfo
  {volume} {77}},\ \bibinfo {pages} {75330} (\bibinfo {year}
  {2008})}\BibitemShut {NoStop}%
\bibitem [{\citenamefont {Liu}\ \emph {et~al.}(2018)\citenamefont {Liu},
  \citenamefont {Guo}, \citenamefont {Vampa}, \citenamefont {Zhang},
  \citenamefont {Sarmiento}, \citenamefont {Xiao}, \citenamefont {Bucksbaum},
  \citenamefont {Vu\'ckovi\'c}, \citenamefont {Fan},\ and\ \citenamefont
  {Reis}}]{Liu2018}%
  \BibitemOpen
  \bibfield  {author} {\bibinfo {author} {\bibfnamefont {H.}~\bibnamefont
  {Liu}}, \bibinfo {author} {\bibfnamefont {C.}~\bibnamefont {Guo}}, \bibinfo
  {author} {\bibfnamefont {G.}~\bibnamefont {Vampa}}, \bibinfo {author}
  {\bibfnamefont {J.}~\bibnamefont {Zhang}}, \bibinfo {author} {\bibfnamefont
  {T.}~\bibnamefont {Sarmiento}}, \bibinfo {author} {\bibfnamefont
  {M.}~\bibnamefont {Xiao}}, \bibinfo {author} {\bibfnamefont {P.}~\bibnamefont
  {Bucksbaum}}, \bibinfo {author} {\bibfnamefont {J.}~\bibnamefont
  {Vu\'ckovi\'c}}, \bibinfo {author} {\bibfnamefont {S.}~\bibnamefont {Fan}},\
  and\ \bibinfo {author} {\bibfnamefont {D.}~\bibnamefont {Reis}},\ }\bibfield
  {title} {\bibinfo {title} {Enhanced high-harmonic generation from an
  all-dielectric metasurface},\ }\href@noop {} {\bibfield  {journal} {\bibinfo
  {journal} {Nature Physics}\ }\textbf {\bibinfo {volume} {14}},\ \bibinfo
  {pages} {1006} (\bibinfo {year} {2018})}\BibitemShut {NoStop}%
\bibitem [{\citenamefont {Ghimire}\ and\ \citenamefont
  {Reis}(2019)}]{Ghimire2019}%
  \BibitemOpen
  \bibfield  {author} {\bibinfo {author} {\bibfnamefont {S.}~\bibnamefont
  {Ghimire}}\ and\ \bibinfo {author} {\bibfnamefont {D.}~\bibnamefont {Reis}},\
  }\bibfield  {title} {\bibinfo {title} {High-harmonic generation from
  solids},\ }\href@noop {} {\bibfield  {journal} {\bibinfo  {journal} {Nature
  Physics}\ }\textbf {\bibinfo {volume} {15}},\ \bibinfo {pages} {10} (\bibinfo
  {year} {2019})}\BibitemShut {NoStop}%
\bibitem [{\citenamefont {Shcherbakov}\ \emph {et~al.}(2021)\citenamefont
  {Shcherbakov}, \citenamefont {Zhang}, \citenamefont {Tripepi}, \citenamefont
  {Sartorello}, \citenamefont {Talisa}, \citenamefont {AlShafey}, \citenamefont
  {Fan}, \citenamefont {Twardowski}, \citenamefont {Krivitsky}, \citenamefont
  {Kuznetsov}, \citenamefont {Chowdhury},\ and\ \citenamefont
  {Shvets}}]{Shcherbakov2021}%
  \BibitemOpen
  \bibfield  {author} {\bibinfo {author} {\bibfnamefont {M.}~\bibnamefont
  {Shcherbakov}}, \bibinfo {author} {\bibfnamefont {H.}~\bibnamefont {Zhang}},
  \bibinfo {author} {\bibfnamefont {M.}~\bibnamefont {Tripepi}}, \bibinfo
  {author} {\bibfnamefont {G.}~\bibnamefont {Sartorello}}, \bibinfo {author}
  {\bibfnamefont {N.}~\bibnamefont {Talisa}}, \bibinfo {author} {\bibfnamefont
  {A.}~\bibnamefont {AlShafey}}, \bibinfo {author} {\bibfnamefont
  {Z.}~\bibnamefont {Fan}}, \bibinfo {author} {\bibfnamefont {Z.}~\bibnamefont
  {Twardowski}}, \bibinfo {author} {\bibfnamefont {L.}~\bibnamefont
  {Krivitsky}}, \bibinfo {author} {\bibfnamefont {A.}~\bibnamefont
  {Kuznetsov}}, \bibinfo {author} {\bibfnamefont {E.}~\bibnamefont
  {Chowdhury}},\ and\ \bibinfo {author} {\bibfnamefont {G.}~\bibnamefont
  {Shvets}},\ }\bibfield  {title} {\bibinfo {title} {Generation of even and odd
  high harmonics in resonant metasurfaces using single and multiple
  ultra-intense laser pulses},\ }\href
  {https://doi.org/10.1038/s41467-021-24450-9} {\bibfield  {journal} {\bibinfo
  {journal} {Nature Communications}\ }\textbf {\bibinfo {volume} {12}},\
  \bibinfo {pages} {4185} (\bibinfo {year} {2021})}\BibitemShut {NoStop}%
\bibitem [{\citenamefont {Zograf}\ \emph {et~al.}(2022)\citenamefont {Zograf},
  \citenamefont {Koshelev}, \citenamefont {Zalogina}, \citenamefont {Korolev},
  \citenamefont {Hollinger}, \citenamefont {Choi}, \citenamefont {Zuerch},
  \citenamefont {Spielmann}, \citenamefont {Luther-Davies}, \citenamefont
  {Kartashov}, \citenamefont {Makarov}, \citenamefont {Kruk},\ and\
  \citenamefont {Kivshar}}]{Zograf2022}%
  \BibitemOpen
  \bibfield  {author} {\bibinfo {author} {\bibfnamefont {G.}~\bibnamefont
  {Zograf}}, \bibinfo {author} {\bibfnamefont {K.}~\bibnamefont {Koshelev}},
  \bibinfo {author} {\bibfnamefont {A.}~\bibnamefont {Zalogina}}, \bibinfo
  {author} {\bibfnamefont {V.}~\bibnamefont {Korolev}}, \bibinfo {author}
  {\bibfnamefont {R.}~\bibnamefont {Hollinger}}, \bibinfo {author}
  {\bibfnamefont {D.-Y.}\ \bibnamefont {Choi}}, \bibinfo {author}
  {\bibfnamefont {M.}~\bibnamefont {Zuerch}}, \bibinfo {author} {\bibfnamefont
  {C.}~\bibnamefont {Spielmann}}, \bibinfo {author} {\bibfnamefont
  {B.}~\bibnamefont {Luther-Davies}}, \bibinfo {author} {\bibfnamefont
  {D.}~\bibnamefont {Kartashov}}, \bibinfo {author} {\bibfnamefont
  {S.}~\bibnamefont {Makarov}}, \bibinfo {author} {\bibfnamefont
  {S.}~\bibnamefont {Kruk}},\ and\ \bibinfo {author} {\bibfnamefont
  {Y.}~\bibnamefont {Kivshar}},\ }\bibfield  {title} {\bibinfo {title}
  {High-harmonic generation from resonant dielectric metasurfaces empowered by
  bound states in the continuum},\ }\href
  {https://doi.org/10.1021/ACSPHOTONICS.1C01511/ASSET/IMAGES/MEDIUM/PH1C01511_M001.GIF}
  {\bibfield  {journal} {\bibinfo  {journal} {ACS Photonics}\ }\textbf
  {\bibinfo {volume} {9}},\ \bibinfo {pages} {567} (\bibinfo {year}
  {2022})}\BibitemShut {NoStop}%
\bibitem [{\citenamefont {Zalogina}\ \emph {et~al.}(2023)\citenamefont
  {Zalogina}, \citenamefont {Carletti}, \citenamefont {Rudenko}, \citenamefont
  {Moloney}, \citenamefont {Tripathi}, \citenamefont {Lee}, \citenamefont
  {Shadrivov}, \citenamefont {Park}, \citenamefont {Kivshar},\ and\
  \citenamefont {Kruk}}]{Zalogina2023}%
  \BibitemOpen
  \bibfield  {author} {\bibinfo {author} {\bibfnamefont {A.}~\bibnamefont
  {Zalogina}}, \bibinfo {author} {\bibfnamefont {L.}~\bibnamefont {Carletti}},
  \bibinfo {author} {\bibfnamefont {A.}~\bibnamefont {Rudenko}}, \bibinfo
  {author} {\bibfnamefont {J.}~\bibnamefont {Moloney}}, \bibinfo {author}
  {\bibfnamefont {A.}~\bibnamefont {Tripathi}}, \bibinfo {author}
  {\bibfnamefont {H.}~\bibnamefont {Lee}}, \bibinfo {author} {\bibfnamefont
  {I.}~\bibnamefont {Shadrivov}}, \bibinfo {author} {\bibfnamefont
  {H.}~\bibnamefont {Park}}, \bibinfo {author} {\bibfnamefont {Y.}~\bibnamefont
  {Kivshar}},\ and\ \bibinfo {author} {\bibfnamefont {S.}~\bibnamefont
  {Kruk}},\ }\bibfield  {title} {\bibinfo {title} {High-harmonic generation
  from a subwavelength dielectric resonator},\ }\href
  {https://doi.org/10.1126/SCIADV.ADG2655/SUPPL_FILE/SCIADV.ADG2655_SM.PDF}
  {\bibfield  {journal} {\bibinfo  {journal} {Science Advances}\ }\textbf
  {\bibinfo {volume} {9}},\ \bibinfo {pages} {eadg2655} (\bibinfo {year}
  {2023})}\BibitemShut {NoStop}%
\bibitem [{\citenamefont {Melik-Gaykazyan}\ \emph {et~al.}(2018)\citenamefont
  {Melik-Gaykazyan}, \citenamefont {Kruk}, \citenamefont {Camacho-Morales},
  \citenamefont {Xu}, \citenamefont {Rahmani}, \citenamefont {Kamali},
  \citenamefont {Lamprianidis}, \citenamefont {Miroshnichenko}, \citenamefont
  {Fedyanin}, \citenamefont {Neshev},\ and\ \citenamefont
  {Kivshar}}]{Liza2018}%
  \BibitemOpen
  \bibfield  {author} {\bibinfo {author} {\bibfnamefont {E.}~\bibnamefont
  {Melik-Gaykazyan}}, \bibinfo {author} {\bibfnamefont {S.}~\bibnamefont
  {Kruk}}, \bibinfo {author} {\bibfnamefont {R.}~\bibnamefont
  {Camacho-Morales}}, \bibinfo {author} {\bibfnamefont {L.}~\bibnamefont {Xu}},
  \bibinfo {author} {\bibfnamefont {M.}~\bibnamefont {Rahmani}}, \bibinfo
  {author} {\bibfnamefont {K.}~\bibnamefont {Kamali}}, \bibinfo {author}
  {\bibfnamefont {A.}~\bibnamefont {Lamprianidis}}, \bibinfo {author}
  {\bibfnamefont {A.}~\bibnamefont {Miroshnichenko}}, \bibinfo {author}
  {\bibfnamefont {A.}~\bibnamefont {Fedyanin}}, \bibinfo {author}
  {\bibfnamefont {D.}~\bibnamefont {Neshev}},\ and\ \bibinfo {author}
  {\bibfnamefont {Y.}~\bibnamefont {Kivshar}},\ }\bibfield  {title} {\bibinfo
  {title} {Selective third-harmonic generation by structured light in
  mie-resonant nanoparticles},\ }\href
  {https://doi.org/10.1021/acsphotonics.7b01277} {\bibfield  {journal}
  {\bibinfo  {journal} {ACS Photonics}\ }\textbf {\bibinfo {volume} {5}},\
  \bibinfo {pages} {728–733} (\bibinfo {year} {2018})}\BibitemShut {NoStop}%
\bibitem [{\citenamefont {Celebrano}\ \emph {et~al.}(2015)\citenamefont
  {Celebrano}, \citenamefont {Wu}, \citenamefont {Baselli}, \citenamefont
  {Grossmann}, \citenamefont {Biagioni}, \citenamefont {Locatelli},
  \citenamefont {De~Angelis}, \citenamefont {Cerullo}, \citenamefont
  {Osellame}, \citenamefont {Hecht}, \citenamefont {Du\'o}, \citenamefont
  {Ciccacci},\ and\ \citenamefont {Finazzi}}]{Celebrano2015}%
  \BibitemOpen
  \bibfield  {author} {\bibinfo {author} {\bibfnamefont {M.}~\bibnamefont
  {Celebrano}}, \bibinfo {author} {\bibfnamefont {X.}~\bibnamefont {Wu}},
  \bibinfo {author} {\bibfnamefont {M.}~\bibnamefont {Baselli}}, \bibinfo
  {author} {\bibfnamefont {S.}~\bibnamefont {Grossmann}}, \bibinfo {author}
  {\bibfnamefont {P.}~\bibnamefont {Biagioni}}, \bibinfo {author}
  {\bibfnamefont {A.}~\bibnamefont {Locatelli}}, \bibinfo {author}
  {\bibfnamefont {C.}~\bibnamefont {De~Angelis}}, \bibinfo {author}
  {\bibfnamefont {G.}~\bibnamefont {Cerullo}}, \bibinfo {author} {\bibfnamefont
  {R.}~\bibnamefont {Osellame}}, \bibinfo {author} {\bibfnamefont
  {B.}~\bibnamefont {Hecht}}, \bibinfo {author} {\bibfnamefont
  {L.}~\bibnamefont {Du\'o}}, \bibinfo {author} {\bibfnamefont
  {F.}~\bibnamefont {Ciccacci}},\ and\ \bibinfo {author} {\bibfnamefont
  {M.}~\bibnamefont {Finazzi}},\ }\bibfield  {title} {\bibinfo {title} {Mode
  matching in multiresonant plasmonic nanoantennas for enhanced second harmonic
  generation},\ }\href {https://doi.org/10.1038/nnano.2015.69} {\bibfield
  {journal} {\bibinfo  {journal} {Nature Nanotechnology}\ }\textbf {\bibinfo
  {volume} {10}},\ \bibinfo {pages} {412} (\bibinfo {year} {2015})}\BibitemShut
  {NoStop}%
\bibitem [{\citenamefont {Frizyuk}\ \emph {et~al.}(2019)\citenamefont
  {Frizyuk}, \citenamefont {Volkovskaya}, \citenamefont {Smirnova},
  \citenamefont {Poddubny},\ and\ \citenamefont {Petrov}}]{frizyuk2019second}%
  \BibitemOpen
  \bibfield  {author} {\bibinfo {author} {\bibfnamefont {K.}~\bibnamefont
  {Frizyuk}}, \bibinfo {author} {\bibfnamefont {I.}~\bibnamefont
  {Volkovskaya}}, \bibinfo {author} {\bibfnamefont {D.}~\bibnamefont
  {Smirnova}}, \bibinfo {author} {\bibfnamefont {A.}~\bibnamefont {Poddubny}},\
  and\ \bibinfo {author} {\bibfnamefont {M.}~\bibnamefont {Petrov}},\
  }\bibfield  {title} {\bibinfo {title} {Second-harmonic generation in
  \uppercase{M}ie-resonant dielectric nanoparticles made of noncentrosymmetric
  materials},\ }\href@noop {} {\bibfield  {journal} {\bibinfo  {journal}
  {Physical Review B}\ }\textbf {\bibinfo {volume} {99}},\ \bibinfo {pages}
  {075425} (\bibinfo {year} {2019})}\BibitemShut {NoStop}%
\bibitem [{\citenamefont {Koshelev}\ \emph {et~al.}(2019)\citenamefont
  {Koshelev}, \citenamefont {Tang}, \citenamefont {Li}, \citenamefont {Choi},
  \citenamefont {Li},\ and\ \citenamefont {Kivshar}}]{Koshelev2019}%
  \BibitemOpen
  \bibfield  {author} {\bibinfo {author} {\bibfnamefont {K.}~\bibnamefont
  {Koshelev}}, \bibinfo {author} {\bibfnamefont {Y.}~\bibnamefont {Tang}},
  \bibinfo {author} {\bibfnamefont {K.}~\bibnamefont {Li}}, \bibinfo {author}
  {\bibfnamefont {D.}~\bibnamefont {Choi}}, \bibinfo {author} {\bibfnamefont
  {G.}~\bibnamefont {Li}},\ and\ \bibinfo {author} {\bibfnamefont
  {Y.}~\bibnamefont {Kivshar}},\ }\bibfield  {title} {\bibinfo {title}
  {Nonlinear metasurfaces governed by bound states in the continuum},\ }\href
  {https://doi.org/10.1021/acsphotonics.9b00700} {\bibfield  {journal}
  {\bibinfo  {journal} {ACS Photonics}\ }\textbf {\bibinfo {volume} {6}},\
  \bibinfo {pages} {1639} (\bibinfo {year} {2019})}\BibitemShut {NoStop}%
\bibitem [{\citenamefont {Tittl}\ \emph {et~al.}(2018)\citenamefont {Tittl},
  \citenamefont {Leitis}, \citenamefont {Liu}, \citenamefont {Yesilkoy},
  \citenamefont {Choi}, \citenamefont {Neshev}, \citenamefont {Kivshar},\ and\
  \citenamefont {Altug}}]{Tittl2018}%
  \BibitemOpen
  \bibfield  {author} {\bibinfo {author} {\bibfnamefont {A.}~\bibnamefont
  {Tittl}}, \bibinfo {author} {\bibfnamefont {A.}~\bibnamefont {Leitis}},
  \bibinfo {author} {\bibfnamefont {M.}~\bibnamefont {Liu}}, \bibinfo {author}
  {\bibfnamefont {F.}~\bibnamefont {Yesilkoy}}, \bibinfo {author}
  {\bibfnamefont {D.-Y.}\ \bibnamefont {Choi}}, \bibinfo {author}
  {\bibfnamefont {D.}~\bibnamefont {Neshev}}, \bibinfo {author} {\bibfnamefont
  {Y.}~\bibnamefont {Kivshar}},\ and\ \bibinfo {author} {\bibfnamefont
  {H.}~\bibnamefont {Altug}},\ }\bibfield  {title} {\bibinfo {title}
  {Imaging-based molecular barcoding with pixelated dielectric metasurfaces.},\
  }\href {https://doi.org/10.1126/science.aas9768} {\bibfield  {journal}
  {\bibinfo  {journal} {Science}\ }\textbf {\bibinfo {volume} {360}},\ \bibinfo
  {pages} {1105} (\bibinfo {year} {2018})}\BibitemShut {NoStop}%
\bibitem [{\citenamefont {Leitis}\ \emph {et~al.}(2019)\citenamefont {Leitis},
  \citenamefont {Tittl}, \citenamefont {Liu}, \citenamefont {Lee},
  \citenamefont {Gu}, \citenamefont {Kivshar},\ and\ \citenamefont
  {Altug}}]{Leitis2019}%
  \BibitemOpen
  \bibfield  {author} {\bibinfo {author} {\bibfnamefont {A.}~\bibnamefont
  {Leitis}}, \bibinfo {author} {\bibfnamefont {A.}~\bibnamefont {Tittl}},
  \bibinfo {author} {\bibfnamefont {M.}~\bibnamefont {Liu}}, \bibinfo {author}
  {\bibfnamefont {B.}~\bibnamefont {Lee}}, \bibinfo {author} {\bibfnamefont
  {M.}~\bibnamefont {Gu}}, \bibinfo {author} {\bibfnamefont {Y.}~\bibnamefont
  {Kivshar}},\ and\ \bibinfo {author} {\bibfnamefont {H.}~\bibnamefont
  {Altug}},\ }\bibfield  {title} {\bibinfo {title} {Angle-multiplexed
  all-dielectric metasurfaces for broadband molecular fingerprint retrieval},\
  }\href {https://doi.org/10.1126/sciadv.aaw2871} {\bibfield  {journal}
  {\bibinfo  {journal} {Science Advances}\ }\textbf {\bibinfo {volume} {5}},\
  \bibinfo {pages} {eaaw2871} (\bibinfo {year} {2019})}\BibitemShut {NoStop}%
\bibitem [{\citenamefont {Xu}\ \emph {et~al.}(2019)\citenamefont {Xu},
  \citenamefont {Shi}, \citenamefont {Davis}, \citenamefont {Yin},\ and\
  \citenamefont {Sievenpiper}}]{shi2019}%
  \BibitemOpen
  \bibfield  {author} {\bibinfo {author} {\bibfnamefont {Z.}~\bibnamefont
  {Xu}}, \bibinfo {author} {\bibfnamefont {J.}~\bibnamefont {Shi}}, \bibinfo
  {author} {\bibfnamefont {R.}~\bibnamefont {Davis}}, \bibinfo {author}
  {\bibfnamefont {X.}~\bibnamefont {Yin}},\ and\ \bibinfo {author}
  {\bibfnamefont {D.}~\bibnamefont {Sievenpiper}},\ }\bibfield  {title}
  {\bibinfo {title} {Rainbow trapping with long oscillation lifetimes in
  gradient magnetoinductive metasurfaces},\ }\href
  {https://doi.org/10.1103/PhysRevApplied.12.024043} {\bibfield  {journal}
  {\bibinfo  {journal} {Phys. Rev. Appl.}\ }\textbf {\bibinfo {volume} {12}},\
  \bibinfo {pages} {024043} (\bibinfo {year} {2019})}\BibitemShut {NoStop}%
\bibitem [{\citenamefont {Tsakmakidis}\ \emph {et~al.}(2007)\citenamefont
  {Tsakmakidis}, \citenamefont {Boardman},\ and\ \citenamefont
  {Hess}}]{boardman2007}%
  \BibitemOpen
  \bibfield  {author} {\bibinfo {author} {\bibfnamefont {K.}~\bibnamefont
  {Tsakmakidis}}, \bibinfo {author} {\bibfnamefont {A.}~\bibnamefont
  {Boardman}},\ and\ \bibinfo {author} {\bibfnamefont {O.}~\bibnamefont
  {Hess}},\ }\bibfield  {title} {\bibinfo {title} {Trapped rainbow storage of
  light in metamaterials},\ }\href@noop {} {\bibfield  {journal} {\bibinfo
  {journal} {Nature}\ }\textbf {\bibinfo {volume} {450}},\ \bibinfo {pages}
  {397} (\bibinfo {year} {2007})}\BibitemShut {NoStop}%
\bibitem [{\citenamefont {Lee}\ \emph {et~al.}(2015)\citenamefont {Lee},
  \citenamefont {Choi}, \citenamefont {Son}, \citenamefont {Park},
  \citenamefont {Ahn},\ and\ \citenamefont {Min}}]{choi2015}%
  \BibitemOpen
  \bibfield  {author} {\bibinfo {author} {\bibfnamefont {K.}~\bibnamefont
  {Lee}}, \bibinfo {author} {\bibfnamefont {H.}~\bibnamefont {Choi}}, \bibinfo
  {author} {\bibfnamefont {J.}~\bibnamefont {Son}}, \bibinfo {author}
  {\bibfnamefont {H.-S.}\ \bibnamefont {Park}}, \bibinfo {author}
  {\bibfnamefont {J.}~\bibnamefont {Ahn}},\ and\ \bibinfo {author}
  {\bibfnamefont {B.}~\bibnamefont {Min}},\ }\bibfield  {title} {\bibinfo
  {title} {Thz near-field spectral encoding imaging using a rainbow
  metasurface},\ }\href {https://doi.org/10.1038/srep14403} {\bibfield
  {journal} {\bibinfo  {journal} {Scientific Reports}\ }\textbf {\bibinfo
  {volume} {5}},\ \bibinfo {pages} {14403} (\bibinfo {year}
  {2015})}\BibitemShut {NoStop}%
\bibitem [{\citenamefont {Stockman}(2004)}]{stockman2004}%
  \BibitemOpen
  \bibfield  {author} {\bibinfo {author} {\bibfnamefont {M.}~\bibnamefont
  {Stockman}},\ }\bibfield  {title} {\bibinfo {title} {Nanofocusing of optical
  energy in tapered plasmonic waveguides},\ }\href@noop {} {\bibfield
  {journal} {\bibinfo  {journal} {Physical Review Letters}\ }\textbf {\bibinfo
  {volume} {93}},\ \bibinfo {pages} {137404} (\bibinfo {year}
  {2004})}\BibitemShut {NoStop}%
\bibitem [{\citenamefont {Davoyan}\ \emph {et~al.}(2010)\citenamefont
  {Davoyan}, \citenamefont {Shadrivov}, \citenamefont {Zharov}, \citenamefont
  {Gramotnev},\ and\ \citenamefont {Kivshar}}]{davoyan2010}%
  \BibitemOpen
  \bibfield  {author} {\bibinfo {author} {\bibfnamefont {A.}~\bibnamefont
  {Davoyan}}, \bibinfo {author} {\bibfnamefont {I.}~\bibnamefont {Shadrivov}},
  \bibinfo {author} {\bibfnamefont {A.}~\bibnamefont {Zharov}}, \bibinfo
  {author} {\bibfnamefont {A.}~\bibnamefont {Gramotnev}},\ and\ \bibinfo
  {author} {\bibfnamefont {Y.}~\bibnamefont {Kivshar}},\ }\bibfield  {title}
  {\bibinfo {title} {Nonlinear nanofocusing in tapered plasmonic waveguides},\
  }\href@noop {} {\bibfield  {journal} {\bibinfo  {journal} {Physical Review
  Letters}\ }\textbf {\bibinfo {volume} {105}},\ \bibinfo {pages} {116804}
  (\bibinfo {year} {2010})}\BibitemShut {NoStop}%
\bibitem [{\citenamefont {Leitis}\ \emph {et~al.}(2021)\citenamefont {Leitis},
  \citenamefont {Tseng}, \citenamefont {Aurelian}, \citenamefont {Kivshar},\
  and\ \citenamefont {Hatice}}]{Leitis2021}%
  \BibitemOpen
  \bibfield  {author} {\bibinfo {author} {\bibfnamefont {A.}~\bibnamefont
  {Leitis}}, \bibinfo {author} {\bibfnamefont {M.}~\bibnamefont {Tseng}},
  \bibinfo {author} {\bibfnamefont {J.-H.}\ \bibnamefont {Aurelian}}, \bibinfo
  {author} {\bibfnamefont {Y.}~\bibnamefont {Kivshar}},\ and\ \bibinfo {author}
  {\bibfnamefont {A.}~\bibnamefont {Hatice}},\ }\bibfield  {title} {\bibinfo
  {title} {Wafer-scale functional metasurfaces for mid-infrared photonics and
  biosensing},\ }\href {https://doi.org/https://doi.org/10.1002/adma.202102232}
  {\bibfield  {journal} {\bibinfo  {journal} {Advanced Materials}\ }\textbf
  {\bibinfo {volume} {33}},\ \bibinfo {pages} {2102232} (\bibinfo {year}
  {2021})}\BibitemShut {NoStop}%
\bibitem [{\citenamefont {Sinev}\ \emph {et~al.}(2021)\citenamefont {Sinev},
  \citenamefont {Koshelev}, \citenamefont {Liu}, \citenamefont {Rudenko},
  \citenamefont {Ladutenko}, \citenamefont {Shcherbakov}, \citenamefont
  {Sadrieva}, \citenamefont {Baranov}, \citenamefont {Itina}, \citenamefont
  {Liu}, \citenamefont {Bogdanov},\ and\ \citenamefont
  {Kivshar}}]{Sinev2021observation}%
  \BibitemOpen
  \bibfield  {author} {\bibinfo {author} {\bibfnamefont {I.}~\bibnamefont
  {Sinev}}, \bibinfo {author} {\bibfnamefont {K.}~\bibnamefont {Koshelev}},
  \bibinfo {author} {\bibfnamefont {Z.}~\bibnamefont {Liu}}, \bibinfo {author}
  {\bibfnamefont {A.}~\bibnamefont {Rudenko}}, \bibinfo {author} {\bibfnamefont
  {K.}~\bibnamefont {Ladutenko}}, \bibinfo {author} {\bibfnamefont
  {A.}~\bibnamefont {Shcherbakov}}, \bibinfo {author} {\bibfnamefont
  {Z.}~\bibnamefont {Sadrieva}}, \bibinfo {author} {\bibfnamefont
  {M.}~\bibnamefont {Baranov}}, \bibinfo {author} {\bibfnamefont
  {T.}~\bibnamefont {Itina}}, \bibinfo {author} {\bibfnamefont
  {J.}~\bibnamefont {Liu}}, \bibinfo {author} {\bibfnamefont {A.}~\bibnamefont
  {Bogdanov}},\ and\ \bibinfo {author} {\bibfnamefont {Y.}~\bibnamefont
  {Kivshar}},\ }\bibfield  {title} {\bibinfo {title} {{Observation of Ultrafast
  Self-Action Effects in Quasi-BIC Resonant Metasurfaces}},\ }\href
  {https://doi.org/10.1021/acs.nanolett.1c03257} {\bibfield  {journal}
  {\bibinfo  {journal} {Nano Lett.}\ }\textbf {\bibinfo {volume} {21}},\
  \bibinfo {pages} {8848} (\bibinfo {year} {2021})}\BibitemShut {NoStop}%
\end{thebibliography}%

\end{document}


\title{Supplementary Information: \\Spectral tuning of high-harmonic generation with resonance-gradient dielectric metasurfaces}

\author{Piyush Jangid}
\altaffiliation{Contributed equally}
\affiliation{Nonlinear Physics Center, Research School of Physics, Australian National University, Canberra ACT 2601, Australia}
\author{Felix Ulrich Richter}
\altaffiliation{Contributed equally}
\affiliation{Institute of Bioengineering, \'Ecole Polytechnique F\'ed\'erale de Lausanne (EPFL), Lausanne 1015, Switzerland}
\author{Ming Lun Tseng}
\affiliation{Institute of Electronics, National Yang Ming Chiao Tung University, Hsinchu 300, Taiwan}
\author{Ivan Sinev}
\affiliation{Institute of Bioengineering, \'Ecole Polytechnique F\'ed\'erale de Lausanne (EPFL), Lausanne 1015, Switzerland}
\author{Sergey Kruk}
\affiliation{Nonlinear Physics Center, Research School of Physics, Australian National University, Canberra ACT 2601, Australia}
\author{Hatice Altug}
\email{Corresponding author: hatice.altug@epfl.ch}
\affiliation{Institute of Bioengineering, \'Ecole Polytechnique F\'ed\'erale de Lausanne (EPFL), Lausanne 1015, Switzerland}
\author{Yuri Kivshar}
\email{Corresponding author: yuri.kivshar@anu.edu.au}
\affiliation{Nonlinear Physics Center, Research School of Physics, Australian National University, Canberra ACT 2601, Australia}



\maketitle  

\section{\label{sup_sec:mode_analysis} Resonator mode analysis}
\begin{figure*} [b]
    \centering
    \includegraphics[width=0.9\textwidth]{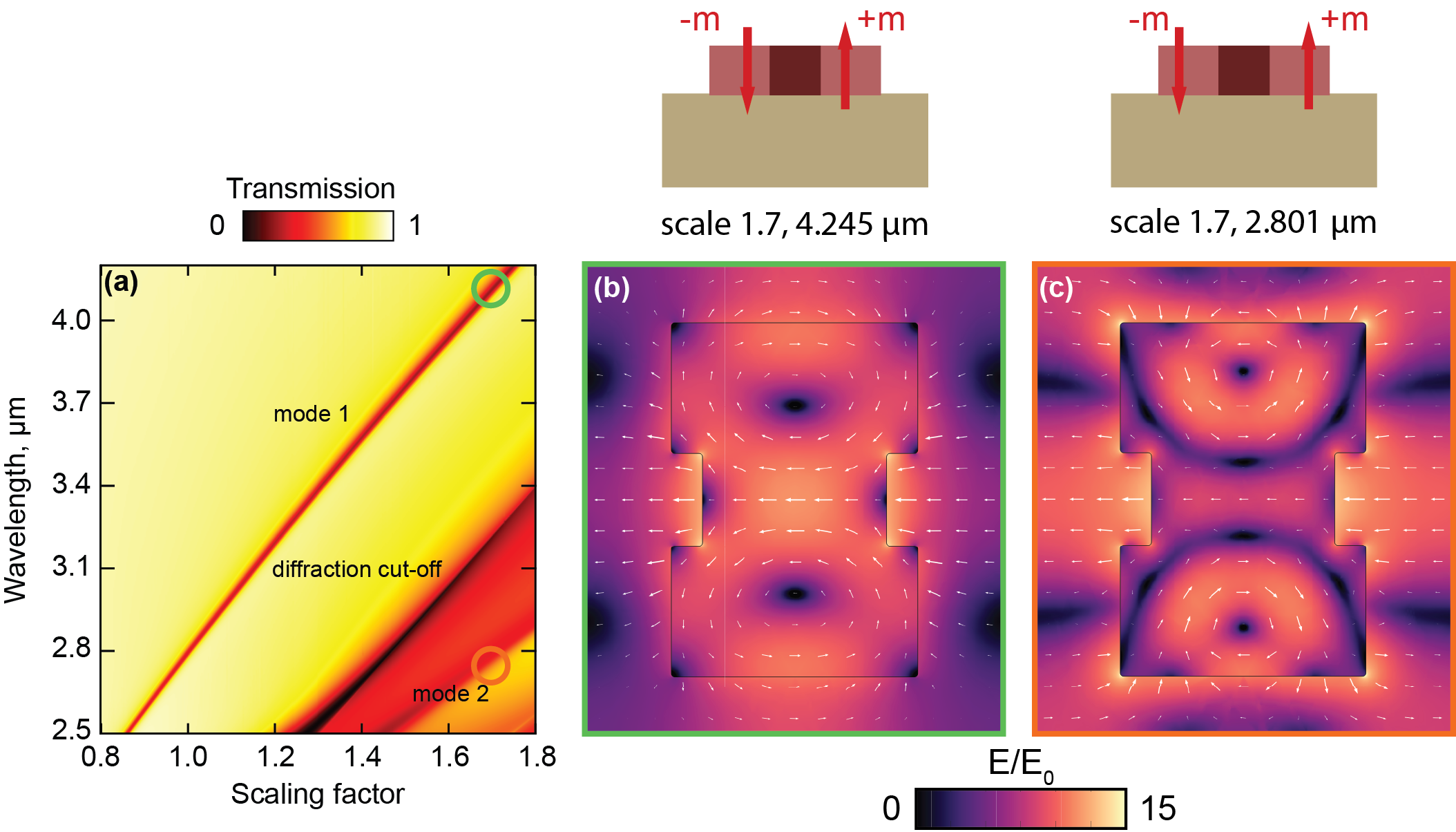}
    \caption{\textbf{Metasurface mode analysis.} \textbf{a.} Calculated transmission map of the metasurface reflectivity depending on the in-plane size scaling parameter. \textbf{b.} Electric field enhancement in the unit cell for the main mode. \textbf{c.} Electric field enhancement in the unit cell for the secondary mode.}
    \label{sup_fig:analysis}
\end{figure*}

As highlighted in the main manuscript, the gradient metasurface exhibits two resonant modes that contribute to the spectrally tunable enhancement of the HHG signal. Fig.~\ref{sup_fig:analysis}a shows the calculated reflectivity map for different scaling factors of the metasurface unit cell for normally incident light polarized along the short axis of the resonator cutout. We choose the scaling factor 1.7 to illustrate the nature of the observed modes. Figs.~\ref{sup_fig:analysis}b and c show the distribution of the electric field enhancement in the unit cell for the wavelengths of 4.245~um and 2.8~um, respectively. The overlayed arrows show the local direction of the in-plane electric field. The field distribution of the main mode demonstrates two characteristic electric field loops that indicate the excitation of two counter-parallel magnetic dipoles in the unit cell at this frequency (illustrated schematically above each of the maps). Similar field distribution is observed for the secondary mode (Fig.~\ref{sup_fig:analysis})c, however, additional e-field minima between the loops indicate that this mode has a higher order than the main one. Lastly, increased reflection in the spectral region between the two modes correspond to the diffraction cut-off.

\begin{figure*} [h]
    \centering
    \includegraphics[width=0.6\textwidth]{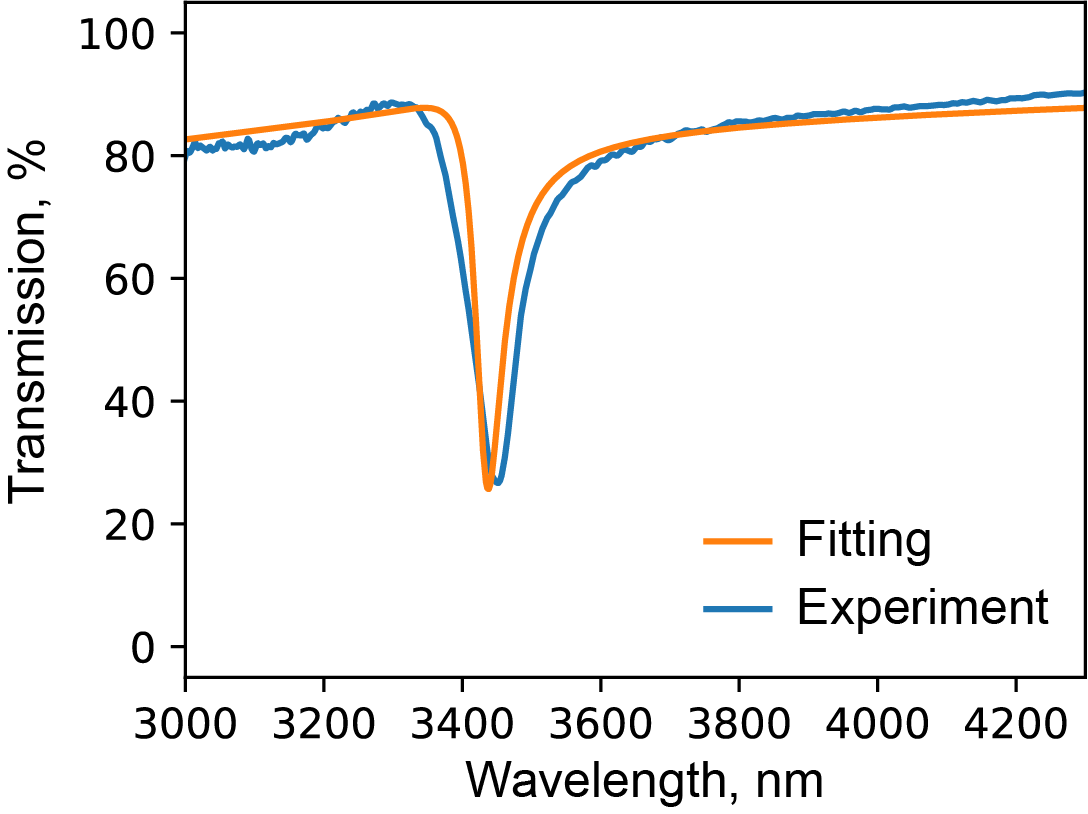}
    \caption{\textbf{Fitting of losses in the simulation.} Experimental metasurface transmission spectrum (blue) and corresponding simulation with artificial Ge loss of 0.018 (orange) show good correspondence both for the resonance linewidth and for the modulation depth.}
    \label{sup_fig:fitting_losses}
\end{figure*}

\section{\label{sup_sec:nanofabrication} Metasurface Fabrication}
The metasurfaces are fabricated on calcium difluoride ($\text{CaF}_2$) substrates featuring minimal absorption losses and low refractive index within the mid-IR spectral range. To enhance the stability of the germanium layer on the $\text{CaF}_2$ substrates, a 5~nm thick layer of silicon oxide ($\text{SiO}_2$) was sputtered as an adhesion layer. Subsequently, a 700~nm layer of germanium was deposited using electron beam evaporation. The desired resonator pattern was defined by utilizing electron beam lithography on spin-coated single-layer PMMA (PMMA 495k A8) films. To facilitate this process, a 25~nm layer of aluminium oxide ($\text{Al}_2\text{O}_3$) serving as hard-mask was deposited through electron beam evaporation and defined by a wet chemical lift-off using commercially available \textit{Remover 1165}. The resonator pattern was then transferred to the underlying germanium layer by fluorine-based dry plasma etching.

\section{\label{sup_sec:efficiency_cal} Efficiency of harmonics generation}

To get the efficiency of generated harmonics, we obtain the average harmonics power by multiplying average intensity counts per spectrometer integration time with the photon energy. We further multiply the average power by a factor $\gamma$, which accounts for the losses through various channels, e.g., via coupling with optical fibre, via spectrometer's diffraction grating, etc. To estimate $\gamma$, we select the pump at the wavelengths of the \3rd and \5th harmonics, and then take the ratio of the average power obtained via powermeter and via the spectrometer.\\

\noindent Below, we calculate the average power of \3rd ($P_{3}$) and \5th ($P_{5}$) harmonics for distinct incident pump powers $P_\omega$ and pump wavelengths $\lambda_{\omega}$.

\begin{table}[h]
\centering
\setlength\extrarowheight{1pt}
\setlength{\tabcolsep}{10pt} 
\begin{tabular}{||c|c||c|c|c|c|c||}
     \hline\hline
     $P_{\omega}$ (mW) &$\lambda_{\omega}$ (nm) &\multicolumn{2}{c|}{$\bar{n}_{3}$ (counts/s)} &$\gamma_3$ &$P_{3}$ (nW) \\[1pt]
     %
     & &\makecell{(Si detector)} &\makecell{(InGaAs detector)} & & \\[1pt]
     \hline\hline
     46 &2526 &$1.794 \times 10^{6}$ & &$3.76 \times 10^4$ &14.59\\[1pt]
     \hline
     46 &2830 &$1.108 \times 10^{6}$ & &$14.84 \times 10^4$ &34.62\\[1pt]
     \hline
     46 &4034 & &$1.855 \times 10^{3}$ &$10.06 \times 10^7$ &27.57\\[1pt]
     \hline\hline
\end{tabular}
\caption{\textbf{\3rd harmonics calculations.} Efficiency = $P_3/P_{\omega}$. $\bar{n}_{3}$ = average \3rd harmonics intensity counts per integration time.}
\label{sup_table:3rd_harmonics}

\vspace{0.8cm}

\begin{tabular}{||c|c||c|c|c|c|c||}
     \hline\hline
     $P_{\omega}$ (mW) &$\lambda_{\omega}$ (nm) &\makecell{$\bar{n}_{5}$ (counts/s) \\(Si detector)} &$\gamma_5$ &$P_{5}$ (pW) \\[1pt]
     \hline\hline
     46 &2830 &$3.620 \times 10^{2}$ &$2.51 \times 10^5$ &31.90\\[1pt]
     \hline
     46 &3250 &$7.703 \times 10^{2}$ &$1.60 \times 10^5$ &37.66\\[1pt]
     \hline
     95.4 &3250 &$3.043 \times 10^{3}$ &$1.60 \times 10^5$ &148.79\\[1pt]
     \hline
     46 &4034 &$1.636 \times 10^{3}$ &$9.37 \times 10^5$ & 37.75\\[1pt]
     \hline\hline
\end{tabular}
\caption{\textbf{\5th harmonics calculations.} Efficiency = $P_5/P_{\omega}$. $\bar{n}_{5}$ = average \5th harmonics intensity counts per integration time.}
\label{sup_table:5th_harmonics}
\end{table}